\documentclass[twocolumn,english,twocolumn,english,aps,prb,floats]{revtex4}
\usepackage[T1]{fontenc}
\usepackage[latin9]{inputenc}
\usepackage{bm}
\usepackage{amsmath}
\usepackage{amssymb}
\usepackage{graphicx}
\usepackage{esint}
\usepackage{adjustbox}
\usepackage{longtable}
\makeatletter
\@ifundefined{textcolor}{}
{%
 \definecolor{BLACK}{gray}{0}
 \definecolor{WHITE}{gray}{1}
 \definecolor{RED}{rgb}{1,0,0}
 \definecolor{GREEN}{rgb}{0,1,0}
 \definecolor{BLUE}{rgb}{0,0,1}
 \definecolor{CYAN}{cmyk}{1,0,0,0}
 \definecolor{MAGENTA}{cmyk}{0,1,0,0}
 \definecolor{YELLOW}{cmyk}{0,0,1,0}
}

\usepackage{babel}

\makeatother

\usepackage{babel}
\begin{document}

\title{Quantum States of a Skyrmion in a 2D Antiferromagnet}

\author{A. Derras-Chouk, E. M. Chudnovsky, and D. A. Garanin}

\affiliation{Physics Department, Herbert H. Lehman College and Graduate School,
The City University of New York, 250 Bedford Park Boulevard West,
Bronx, New York 10468-1589, USA }

\date{\today}
\begin{abstract}
Quantum states of a skyrmion in a 2D antiferromagnetic lattice are obtained by quantizing the scaling parameter of Belavin-Polyakov model. Skyrmion classical collapse due to violation of the translational invariance of the continuous spin-field model by the lattice is replaced in quantum mechanics by transitions between discrete energy levels of the skyrmion. Rates of transitions due to emission of magnons are computed. Ways of detecting  quantization of skyrmion states are discussed.
\end{abstract}

\maketitle

\section{Introduction} \label{Intro}

Skyrmions came to material science \cite{BP} from high-energy physics where they were introduced to model atomic nuclei \cite{SkyrmePRC58,Polyakov-book,Manton-book}. They are prospective candidates for topologically protected magnetic memory \cite{Leonov-NJP2016,Fert-Nature2017,Bogdanov2020}. Topological stability of skyrmions arises from discrete homotopy classes of mapping of the continuous field on the continuous geometrical space, e.g., mapping of a three-component constant-length spin field onto the 2D space of a magnetic film. It relies on the translation (to be exact, conformal) invariance of the 2D Heisenberg model. As soon as this invariance is broken by the crystal lattice, skyrmions become unstable against collapsing \cite{CCG-PRB2012} and must be stabilized by additional interactions, such as Dzyaloshiskii-Moriya, magnetic anisotropy, Zeeman, etc. In a typical experiment the size of the skyrmion is controlled by the magnetic field. Below a certain size the exchange interaction always wins and the skyrmion collapses \cite{Muckel2021}. 

The observed skyrmion textures typically encompass thousands or spins. Even the smallest nanoscale skyrmions experimented with are still comprised of hundreds of spins. Such skyrmions were imaged by the Lorentz transmission electron microscopy \cite{Back2020} and are generally perceived as classical objects. As the skyrmion becomes smaller, however, one must expect that at some point quantum mechanics comes into play. This work is motivated by the observation that classical collapse of a skyrmion into a point of a crystal lattice is at odds with quantum mechanics. It contradicts the uncertainty principle the same way as the classical collapse of an electron onto a proton does. The problem at hand is much more difficult, however, than the problem of the hydrogen atom. Huge number of spin degrees of freedom possessed by the skyrmion resembles the problem of the many-electron atom for which analytical computation of quantum states is impossible. 

Some aspects of the quantum behavior of skyrmions have been addressed in the past. Quantum motion of a skyrmion in the pinning potential has been studied, based upon the analogy of the Thiele dynamics with the motion of a charged particle in the magnetic field \cite{Lin2013}. Magnon-skyrmion scattering in chiral magnets has been addressed by deriving Bolgoliubov - de Gennes Hamiltonian from the Lagrangian of the spin field \cite{Schutte2014}. By developing Holstein-Primakoff transformation of a skyrmion texture, quantum spin excitations of the skyrmion have been obtained \cite{Roldan2015,Oh2015} and it was shown that quantum fluctuations tend to stabilize skyrmion textures. A particle model of a skyrmion quantum liquid emerging from the melting of a skyrmion crystal has been proposed \cite{Takashima-PRB2016}. Quantum tunneling of a skyrmion under the energy barrier created by competing interactions has been studied within semiclassical approach based upon Euclidean action for the spin field \cite{Amel2018,Vlasov2020}. Evidence of quantum skyrmion states has been obtained by exact diagonalization of the Heisenberg Hamiltonian of a frustrated ferromagnet \cite{Lohani-PRX2019}.  A review of quantum skyrmionics highlighting the relation of the problem to Chern-Simons theories and quantum Hall effect has been given in Ref.\ \onlinecite{Ochoa2019}. Recently, quantum computer simulator has been utilized to obtain quantum skyrmion states in a lattice model with Heisenberg, Dzyaloshinskii-Moriya, and Zeeman interactions \cite{Sotnikov-PRB2021}.

In this paper we take a different approach to the quantization of the skyrmion field. A Belavin-Polyakov skyrmion \cite{BP} is characterized by a scaling parameter $\lambda$ that can be roughly interpreted as its size. In a continuous spin-field exchange model the energy of the skyrmion is independent of $\lambda$. However, in a discrete model with a finite lattice spacing $a$ the energy acquires \cite{CCG-PRB2012} a term proportional to $-(a/\lambda)^2$. It leads to the collapse of a classical skyrmion to a point, $\lambda \rightarrow 0$. In quantum mechanics the lattice term can be interpreted as a potential well $U(\lambda)$ inside which the skyrmion must have quantized energy levels. In antiferromagnets, inertia associated with the dynamics of the N\'{e}el vector  allows one to introduce the conjugate momentum associated with the generalized coordinate $\lambda$, thus making quantization of the problem straightforward. It is conceptually similar to the quantization of a string loop collapsing under tension \cite{Klebanov}. 

The article is organized as follows. Classical dynamics of the antiferromagnetic skyrmion in a 2D crystal lattice is discussed in Section \ref{classical}. Quantization of the Hamiltonian of the skyrmion is performed in Section \ref{QH}. Eigenfunctions and eigenvalues of the skyrmion in the potential well created by the lattice are obtained. Rates of the transitions between quantized states of the skyrmion, accompanied by the radiation of a magnon, are computed in Section \ref{Rates}.  Possible systems and experiments are discussed in Section \ref{Discussion}.

\section{Classical skyrmion on a lattice} \label{classical}

We begin with the exchange Hamiltonian of a 2D antiferromagnet in a continuous spin-field approximation,
\begin{equation}
\mathcal{H}_{0} = \frac{1}{2}JS^2\int
dxdy\, \left(\frac{1}{c^2}\partial_{t}\mathbf{L} \cdot \partial_{t}\mathbf{L}+\partial_{i}\mathbf{L} \cdot \partial_{i}\mathbf{L}\right) \label{H0}.
\end{equation}
Here $J > 0$ is a constant of the exchange interaction between nearest-neighbor spins of length $S$,  ${\bf L}$ is a normalized N\'{e}el vector, and $c$ is the speed of antiferromagnetic spin waves that equals $2\sqrt{2}JSa/\hbar$ in a square lattice. Summation over the repeated index $i=x,y$ is assumed. The first term in Eq.\ (\ref{H0}) can be interpreted as the kinetic energy responsible for the inertia of the spin field in antiferromagnets \cite{Haldane,Ivanov,CCG-PRB2012}. In the low-energy domain, strong antiferromagnetic exchange between antiparallel sublattices makes the length of the N\'{e}el vector nearly constant,  ${\bf L}^{2}=1$. Hamiltonian (\ref{H0}) is equivalent to the $\sigma$-model in relativistic field theory \cite{SkyrmePRC58}.

Within the continuous field theory based upon Hamiltonian (\ref{H0}) skyrmions are stable due to the conservation of the topological charge  $Q=\frac{1}{4\pi}\int dxdy\,{\bf L}\cdot(\partial_{x}{\bf L}\times\partial_{y}{\bf L})\label{Q}$ that takes values $Q=0,\pm1,\pm2,\ldots$. Skyrmion with $Q = \pm1$ is given by \cite{BP}
\begin{equation}
{\bf L}=\left(\frac{2\lambda r\cos(\varphi_r + \gamma)}{r^2+\lambda^{2}},\,Q\frac{2\lambda r\sin(\varphi_r + \gamma)}{r^2+\lambda^{2}},\, \frac{r^2-\lambda^{2}}{r^2+\lambda^{2}}\right)\,,
\label{sn}
\end{equation}
where $r$ and $\varphi_r$ are polar coordinates in the 2D plane, $\gamma$ is an arbitrary chirality angle, and $\lambda$ is an arbitrary scaling parameter. Crucial for our treatment of the quantum problem is observation that $\lambda$ can be both positive and negative. Its sign is related to the chirality of the skyrmion while its modulus can be interpreted as the skyrmion size. The energy of the  skyrmion, $E_0=4\pi JS^{2}$, is independent of $\gamma$ and $\lambda$. 

Lattice of a finite spacing $a$ breaks the stability of the skyrmion by making its energy depend on $\lambda$. This dependence was worked out in Ref.\ \onlinecite{CCG-PRB2012} within a Heisenberg model with two antiferromagnetic sublattices in a square lattice. The lattice contribution to the Hamiltonian is
\begin{equation}
\mathcal{H}_{\rm lat} = -\frac{a^2}{24}JS^2\int
dxdy\, \partial_{i}^{2}\mathbf{L}\cdot \partial_{i}^{2}\mathbf{L}\label{Ham-lat}.
\end{equation}
Treating this term as a perturbation and substituting Eq.\ (\ref{sn}) into Eq.\ (\ref{Ham-lat}) one obtains for the energy of the skyrmion with $\lambda \gtrsim a$
\begin{equation}
E = 4\pi JS^{2}\left[1 - \frac{1}{6}\left(\frac{a}{\lambda}\right)^2\right].
\label{EDiscr}
\end{equation}
This result can be generalized \cite{CGC-PRB2019} for an arbitrary $Q$. It is independent of the chirality angle $\gamma$. It shows that due to a nonzero $a$ the skyrmion would decrease its energy by collapsing towards smaller $\lambda$. 

\begin{figure}[h]
\centering{}
\includegraphics[width=9.0cm]{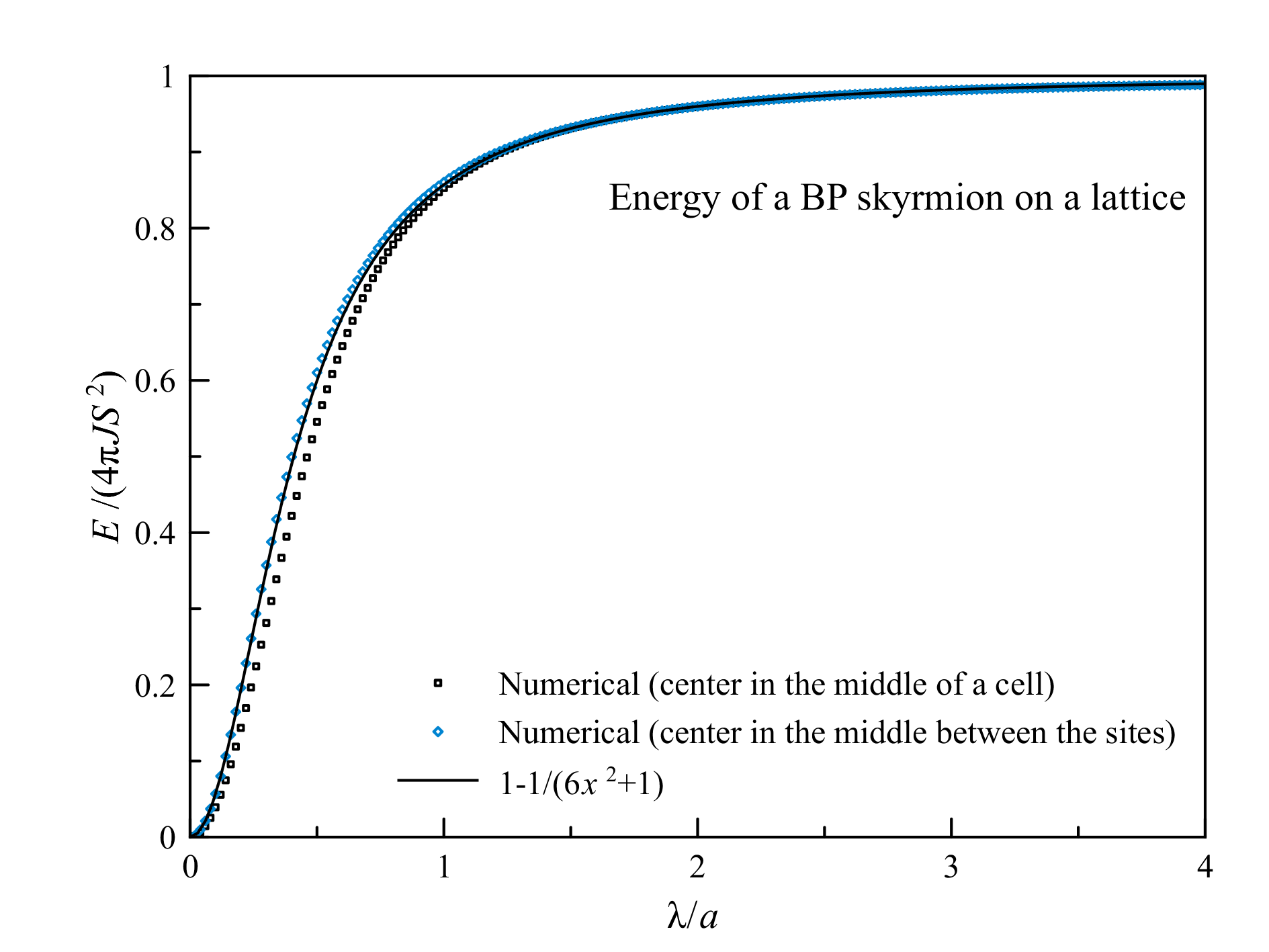} 
\caption{Dependence of the energy of the Belavin-Polyakov skyrmion on its size, computed numerically in a square lattice of spins.}
\label{numerical}
\end{figure}

The dynamics of the collapse is described by the Hamiltonian $\mathcal{H} = \mathcal{H}_{0} + \mathcal{H}_{\rm lat}$.   For the skyrmion given by Eq.\ (\ref{sn}) one has \cite{CCG-PRB2012}
\begin{equation}
{\cal{H}} = E_0 + \frac{\pi \hbar^2}{2J a^2} \ln\left(\frac{l/\sqrt{e}}{\sqrt{\lambda^2 + a^2/6}}\right) \left(\frac{d\lambda}{dt}\right)^2 -\frac{2\pi J S^2 a^2}{3(\lambda^2 + a^2/6)}. \label{Ham}
\end{equation}
Here we have introduced a large-distance cutoff $l$ due to the finite size of the 2D system and a small-distance cutoff due to the discreteness of the crystal lattice. The latter was chosen such as to eliminate the unphysical discontinuities at $\lambda \rightarrow 0$ in the denominators of Eq.\ (\ref{Ham}) and provide the zero static energy at $\lambda = 0$. This choice is supported by the direct numerical summation for the energy on the lattice, using skyrmion profile in Fig.\ (\ref{sn}) with different $\lambda$, see Fig.\ \ref{numerical}. Excellent fit of the microscopic many-spin result by the potential $V = - 4\pi JS^2a^2/(6\lambda^2 + a^2)$ of Eq.\ (\ref{Ham}) allows one to extend it to the region  $\lambda < a$. 

\begin{figure}[h]
\centering{}
\includegraphics[width=8.5cm]{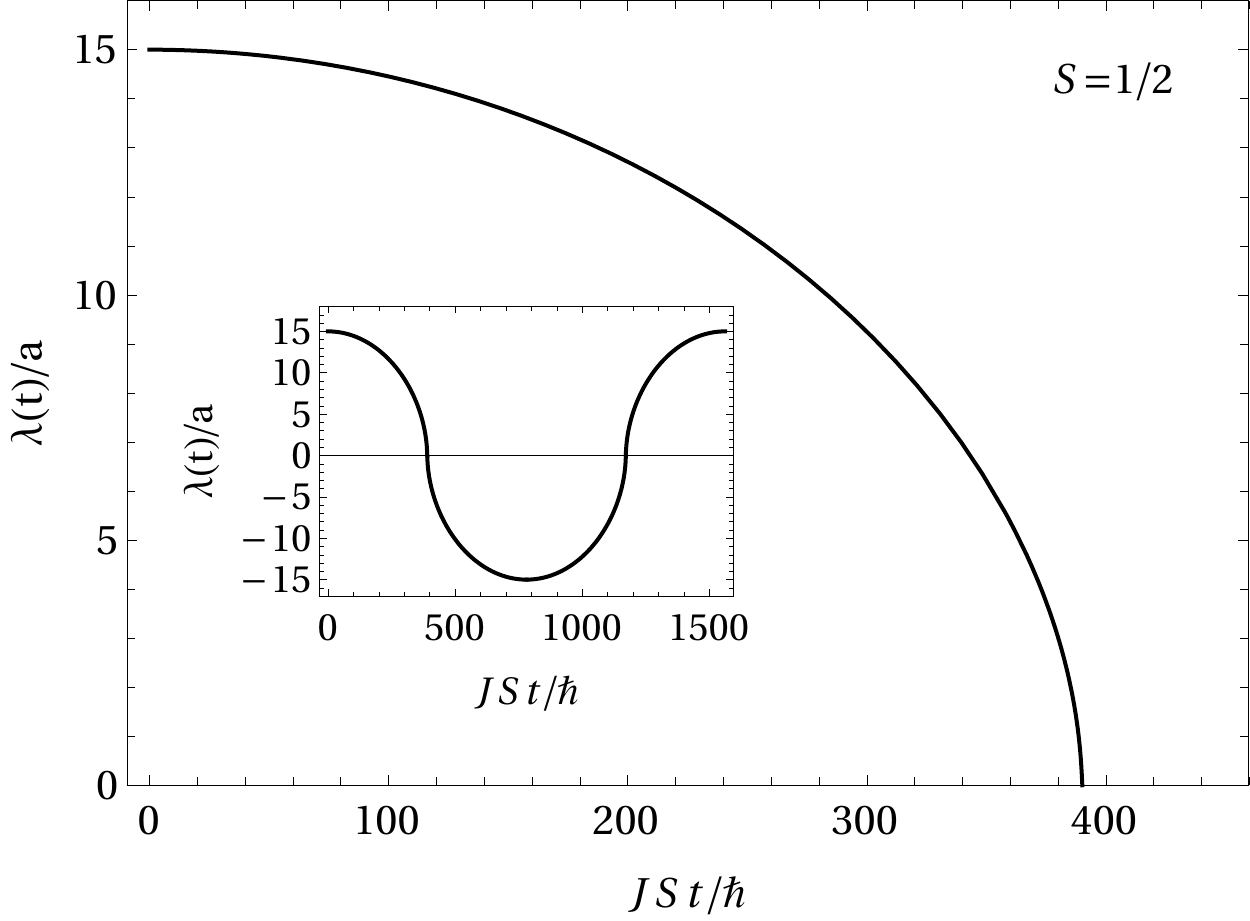} 
\caption{Classical collapse of a skyrmion with from the intial size $\lambda_0 = 15a$ in a circular 2D space of radius $l = 1000a$. The inset shows periodic oscillations of the skyrmion between states of opposite chirality described by Eq.\ \ref{motion}.}
\label{dynamics}
\end{figure}
The behavior of $\lambda(t)$ corresponding to the collapse of a classical skyrmion of the initial size $\lambda_0 = 15a$ in a circular 2D space of radius $l = 1000a$, that follows from the conservation of energy,
\begin{equation}
\left(\frac{dx}{dt}\right)^2\ln\left(\frac{\bar{l}/\sqrt{e}}{\sqrt{x^2 + 1/6}}\right) =   \frac{4}{3}\left[\frac{1}{x^2 + 1/6} -\frac{1}{x_0^2 + 1/6}\right], \label{motion}
\end{equation}
is shown in Fig.\ \ref{dynamics}. It is derived from the Hamiltonian (\ref{Ham}) by choosing $x = \lambda/a$, $\bar{t} = JSt/\hbar$ and the initial state that starts from rest, $dx/dt = 0$, with $x = x_0 = \lambda_0/a$. The dependence of the skyrmion lifetime on the system size $\bar{l} =l/a$ is weak. 

Temporal behavior of the collapsing skyrmion shown in Fig.\ \ref{dynamics} was confirmed by the numerical study of the full two-sublattice classical antiferromagnetic Heisenberg spin model on a square lattice  \cite{CCG-PRB2012}.  That model also captured the decay of the topological charge to zero at the final stage of the collapse. Based upon numerical results it was argued that the collapse of the skyrmion towards lower energies via the reduction of $\lambda$ was accompanied by the radiation of magnons. If this effect was neglected, the skyrmion would collapse and expand in a periodic manner, oscillating in the potential well $V = - 4\pi JS^2a^2/(6\lambda^2 + a^2)$ created by the lattice between positive and negative $\lambda$ corresponding to opposite chiralities, see inset in Fig.\ \ref{dynamics}. 

Here we notice that the classical theory of the skyrmion collapse comes into conflict with quantum mechanics. The contraction of a skyrmion to a point accompanied by its growing radial momentum contradicts the uncertainty principle. Quantum mechanics should suppress continuous radiation of magnons by the skyrmion as it suppresses continuous radiation of electromagnetic waves by a classical electron falling onto a proton. This must make skyrmions more stable against the collapse. In the correct description the skyrmion must have quantized energy levels in the lattice potential and probability distribution of the skymion size. Its expectation value must be computed as
\begin{equation}
 \bar{\lambda} = \sqrt{\langle \lambda^2 \rangle} = \left[\int d\lambda \psi^{*}(\lambda) \lambda^2 \psi(\lambda)\right]^{1/2} \label{rms}
 \end{equation}
based upon the knowledge of the skyrmion wave function $\psi(\lambda)$.

\section{Quantum Hamiltonian} \label{QH}

Hamiltonian (\ref{Ham}) can be viewed as a Hamiltonian of a particle with a coordinate $\lambda$ and a mass
\begin{equation}
M(\lambda) = \frac{\pi\hbar^2}{J a^2} \ln\left(\frac{l/\sqrt{e}}{\sqrt{\lambda^2 + a^2/6}}\right)
\end{equation}
that depends logarithmically on $l$ and $\lambda$. Its typical value for, e.g.,  $J \sim 1000$K and $a \sim 0.3$nm is $M \sim 10^{-28}$kg, which is about one hundred electron masses. Up to a log factor it coincides with the mass of the antiferromagnetic skyrmion, $M = E_0/c^2$, that can be obtained by substituting Eq.\ (\ref{sn}) with $x$ replaced by $x - vt$ into Eq.\ (\ref{H0}). 

\begin{table*}
\begin{centering}
\begin{adjustbox}{width=17.8cm}
\begin{tabular}{|c|c|c|c|c|c|c|c|c|c|c|c|c|c|}
\hline 
$n$ & 0 & 1 & 2 & 3 & 4 & 5 & 6 & 7 & 8 & 9 & 10 & 11 & 12\tabularnewline
\hline 
$\bar{\lambda}_{n}/a$ & 0.1518 & 0.2994 & 0.4663 & 0.6840 & 0.9841 & 1.410 & 2.029 & 2.943 & 4.318 & 6.428 & 9.741 & 15.09 & 24.03\tabularnewline
\hline 
$E_{n}/J$ & -2.581 & -1.635 & -0.9861 & -0.5666 & -0.3108 & -0.1634 & -0.08273 & -0.04043 & -0.01910 & -0.008715 & -0.003827 & -0.001607 & -0.0006391\tabularnewline
\hline 
\end{tabular}
\end{adjustbox}
\par\end{centering}
\caption{Energy levels and rms skyrmion sizes for $S=1/2$.}
\label{table1}
\end{table*}

\begin{figure}[h]
\centering{}
\includegraphics[width=8.6cm]{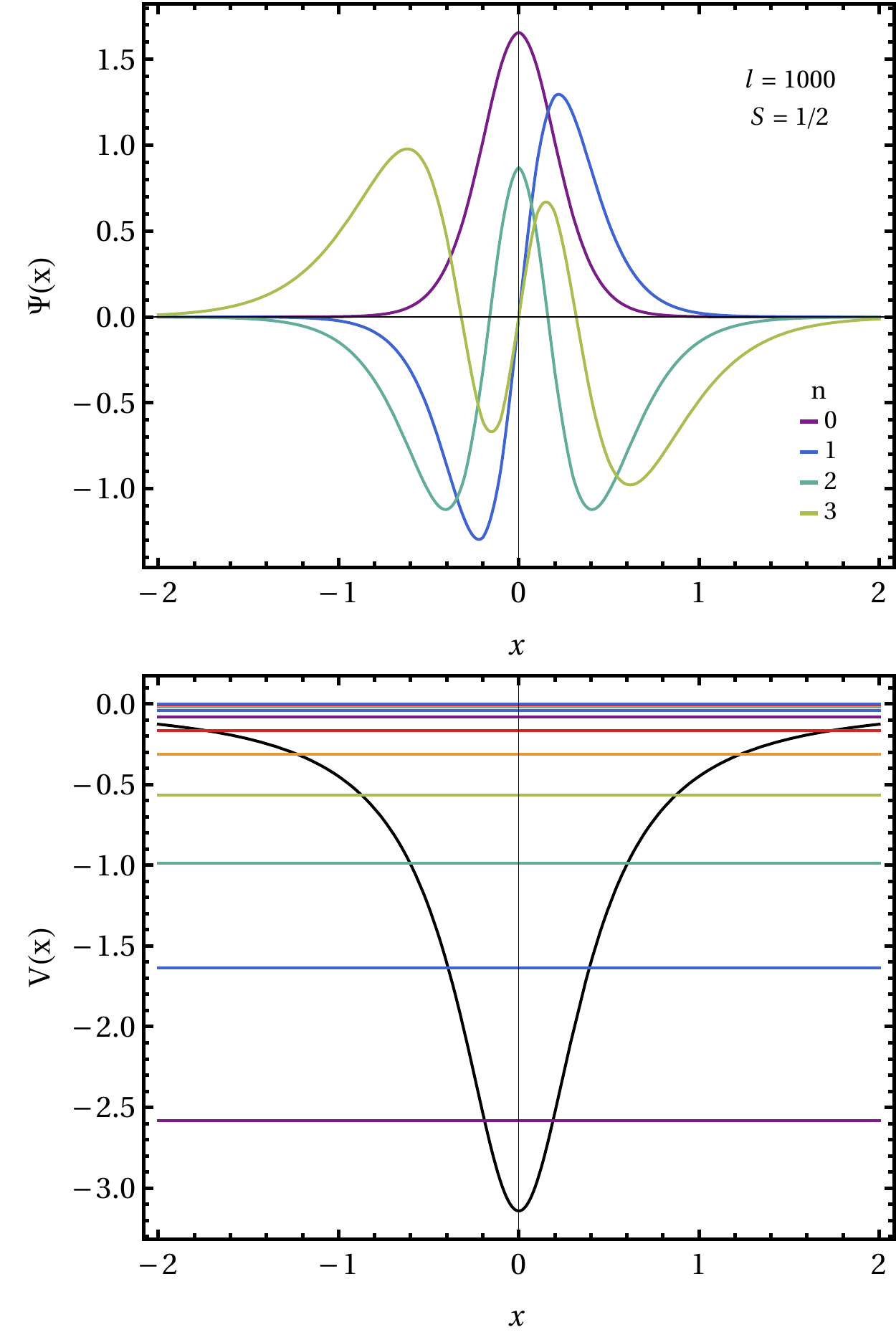} 
\caption{Upper panel: Eigenfunctions of four lowest-energy states for $S = 1/2$, $l = 1000a$ scaled to $x= \lambda/a$. Lower panel:  Energy levels of the skyrmion for S = 1/2 in the units of $J$ in the potential well $V(x)=-4\pi S^2/(6x^2+1)$.}
\label{levels}
\end{figure}
The generalized momentum is $p = M(d\lambda/dt)$, so that up to a constant $E_0$ the Hamiltonian can be written as
\begin{equation}
{\cal{H}} = \frac{p^2}{2M(\lambda)} -\frac{2\pi J S^2 a^2}{3(\lambda^2 + a^2/6)}.
\end{equation}
Imposing quantization as
\begin{equation}
\lambda \hat{p} - \hat{p} \lambda = i\hbar
\end{equation}
and writing
\begin{equation} 
\hat{p} = -i\hbar \frac{d}{d \lambda}
\end{equation}
we obtain 
\begin{eqnarray}
&& {\cal{H}} = \hat{p}\frac{1}{2M(\lambda)}\hat{p} - \frac{2\pi J S^2 a^2}{3(\lambda^2 + a^2)}  \label{Sr} \\
& &  =  -\frac{Ja^2}{2\pi}\frac{d}{d\lambda}\left[{\ln^{-1}\left(\frac{l/\sqrt{e}}{\sqrt{\lambda^2 + a^2/6}}\right)}\frac{d}{d\lambda}\right] - \frac{2\pi J S^2 a^2}{3(\lambda^2 + a^2/6)}. \nonumber
\end{eqnarray}
It is easy to check that this Hamiltonian is Hermitian for symmetric and antisymmetric wave functions of the bound states. Classical limit corresponds to $S \rightarrow \infty$ when the potential energy dominates over the kinetic energy, or to $l \rightarrow \infty$ when the mass of the ``particle'' becomes infinite. In the numerical work presented below we use $S=1/2$ and $l = 1000$. The dependence on $l$ is weak while results for other $S$ are qualitatively similar because choosing a different $S$ only changes the depth of the potential well.

Eigenstates of the Hamiltonian (\ref{Sr}) for the discrete energy spectrum at $S = 1/2$ and $l = 1000a$  are shown in Fig.\ \ref{levels}. Finite element discretization method with Arnoldi algorithm and shooting have been used and compared with each other. The density of energy levels increasing towards the top of the potential well created  by the lattice. These states represent symmetric and antisymmetric quantum oscillations between opposite chiralities corresponding to positive and negative $\lambda$.  

\begin{figure}[h]
\centering{}
\includegraphics[width=8.9cm]{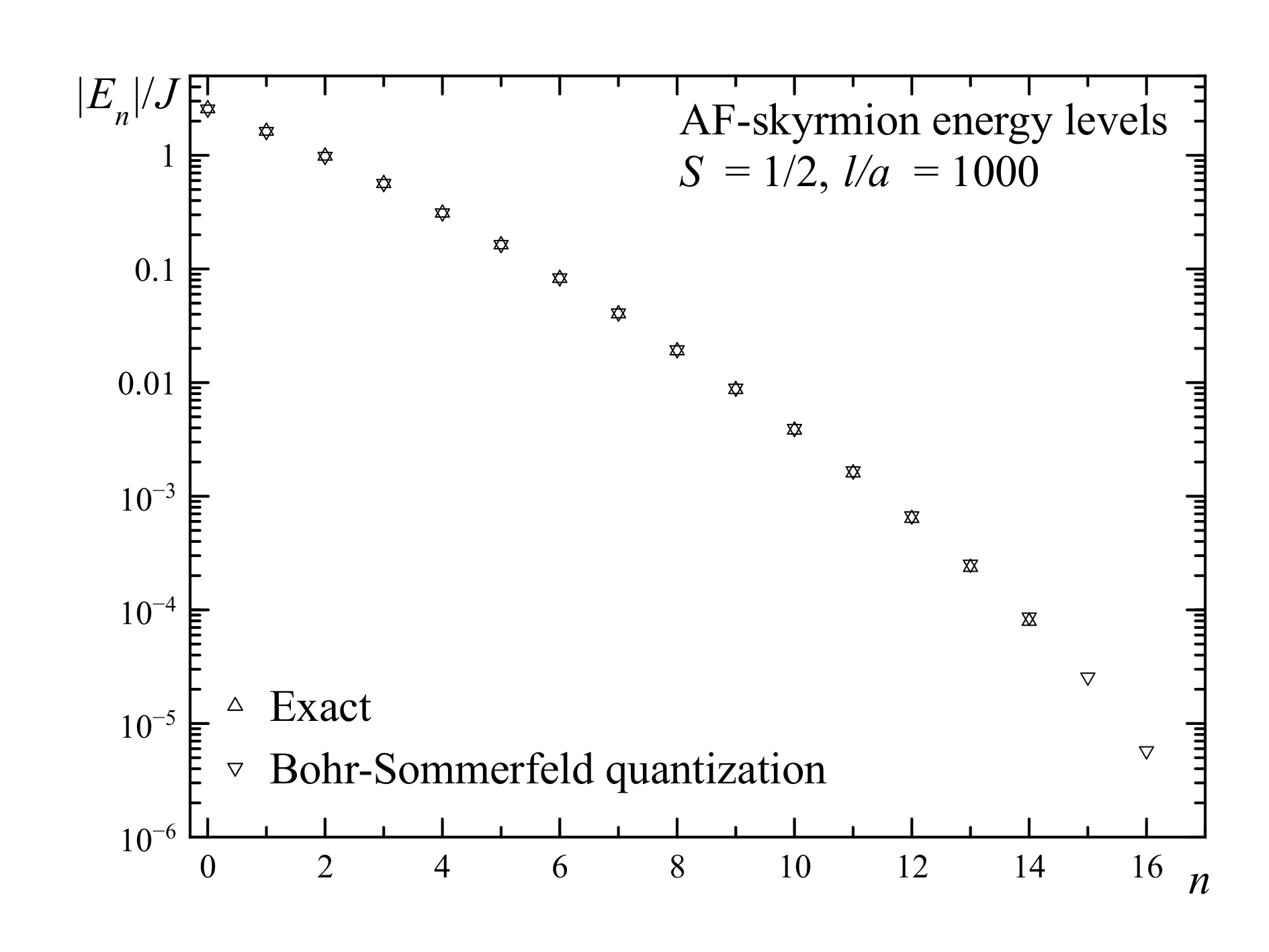}  
\caption{Energy levels computed by solving the Schr$\ddot{\text{o}}$dinger equation for the eigenstates and obtained with the help of the Bohr-Sommerfeld quantization condition, Eq.\ (\ref{BS}).}
\label{Bohr}
\end{figure}
Note that energy eigenstates can be alternatively found from the Bohr-Sommerfeld condition: $\oint d\lambda\, p(\lambda) = 2\pi \hbar (n + 1/2)$, which reduces in our case to solving the equation
\begin{equation}
\int^{x_n}_{-x_n}dx \,\left\{\frac{2}{\pi } \ln\left(\frac{\bar{l}/\sqrt{e}}{\sqrt{x^2 + 1/6}}\right)\left[\bar{E}_n - \bar{V}(x)\right]\right\}^{1/2}  = n +\frac{1}{2} \label{BS}
\end{equation}
with $x = {\lambda}/{a}, \bar{l} = {l}/{a}$,  $\bar{E} = {E}/{J}$, $\bar{V} = V/J = -4\pi S^2/(6x^2 + 1)$, for $\bar{E}_n = \bar{V}(x_n)$, and $n = 0,1,2,...$. Allowed values of $n$ are restricted by the condition $x_n < \bar{l}$. This quasiclassical method works surprisingly well for small $n$, see Fig.\ \ref{Bohr}. The error is $1.3\%$ for the ground state and $-4.9\%$ for $n = 12$ as compared to the values obtained by solving  Schr$\ddot{\text{o}}$dinger equation. For large $n$ the decrease of $E_n$ on increasing $n$ is faster than $\exp(-n)$. 

\begin{figure}[h]
\centering{}
\includegraphics[width=8.9cm]{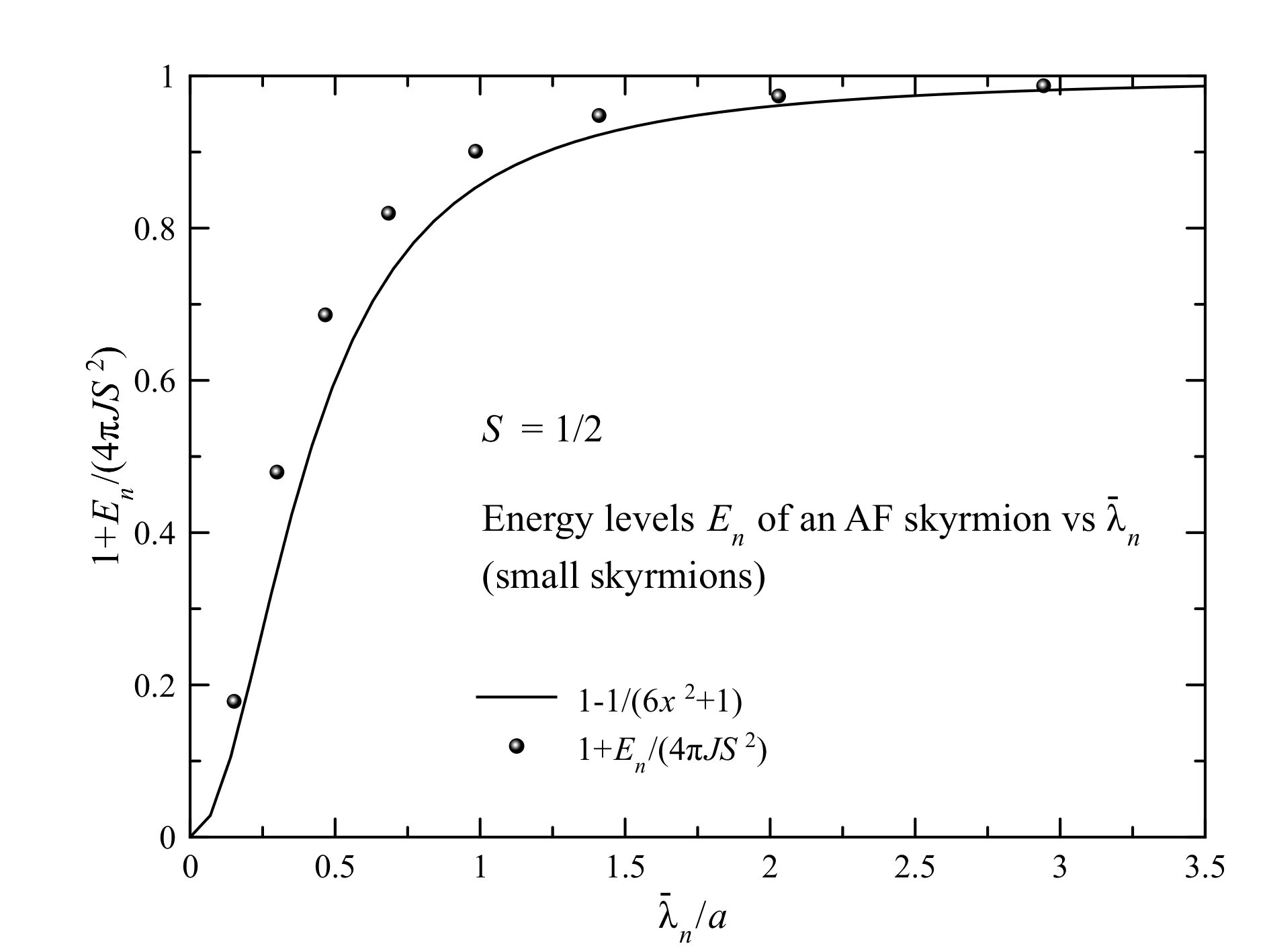} 
\includegraphics[width=8.9cm]{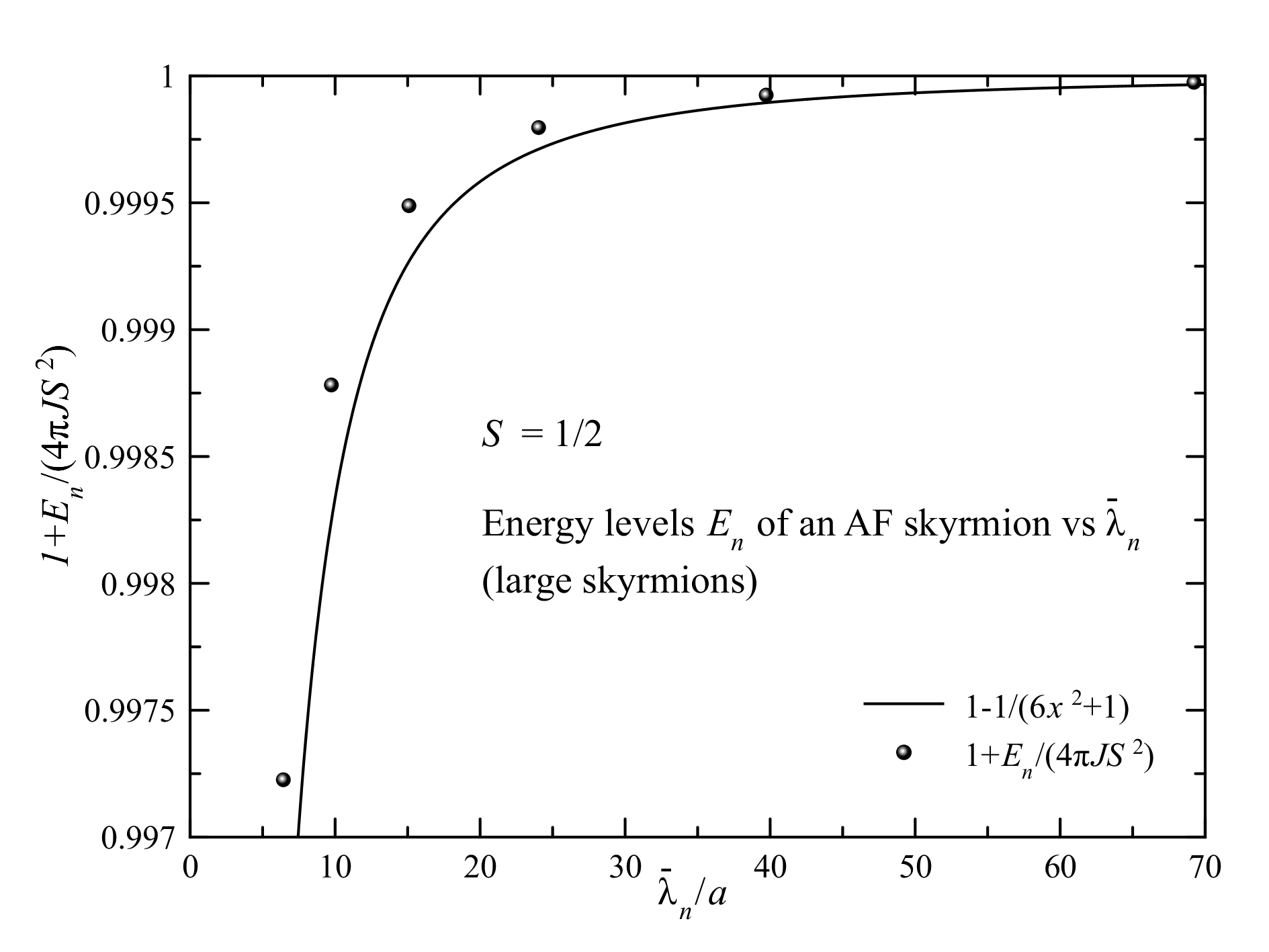} 
\caption{Relation between energies and rms sizes of quantized skyrmion states is determined by the shape of the potential due to the lattice. Upper panel: Small skyrmions with $S = 1/2$ in a lattice with $l = 1000a$. Lower panel: Large skyrmions.}
\label{corr}
\end{figure}

For each of the quantum states one can compute the rms value of $\lambda$ according to Eq.\ (\ref{rms}) with the wave function $\psi(\lambda)$ shown in the upper panel of Fig.\ \ref{levels}. The first twelve $\bar{\lambda}_n$ are listed in Table \ref{table1} together with the corresponding energy levels. An interesting observation is in order. While distances between the energy levels decrease exponentially as one approaches the top of the potential well, the distances between the corresponding skyrmion rms sizes increase. This can be qualitatively understood by showing positions of the energy levels and skyrmion rms sizes together with the attractive potential, $V = - 4\pi JS^2a^2/(6\lambda^2 + a^2)$, see Fig.\ \ref{corr}. Correlation $\bar{E}_n \sim V(\bar{\lambda}_n)$ is apparent from the figure. 
Due to the quantization of skyrmion states the transitions between densely packed energy levels of a collapsing nanoscale skyrmion must occur via sizable jumps towards smaller skyrmion sizes. 

\section{Transitions between quantum states of a skyrmion} \label{Rates}

The quantum counterpart of the continuous skyrmion collapse in a classical theory are quantum transitions from higher to lower energy levels with lower $\bar{\lambda}_n$. They are accompanied by radiation of magnons that correspond to quantized linear waves ${\bf \delta L}({\bf r}, t)$ of ${\bf L}$. Since a skyrmion is an exact extremum of the Hamiltonian (\ref{H0}),  it does not interact with linear excitations to the first order on $\delta {\bf L}$. This is easy to see from the fact that skyrmion solutions are obtained from the variation of the exchange Hamiltonian under the condition ${\bf L}^2 = 1$,
\begin{equation}
\delta\mathcal{H} = -\frac{1}{2}JS^2\int
dxdy\, \left[{\partial_i}^2{\bf L} - ({\bf L}\cdot {\partial_i}^2{\bf L}){\bf L}\right]\cdot \delta{\bf L}, \label{variation}
\end{equation}
which provides the extremal equation for the skyrmion ${\partial_i}^2{\bf L} = ({\bf L}\cdot {\partial_i}^2{\bf L}){\bf L}$. 

Interaction of skyrmions with magnons comes from the lattice term given by Eq.\ (\ref{Ham-lat}). Writing ${\bf L}$ in that term as ${\bf L}_s + \delta {\bf L}$, where ${\bf L}_s$ is the skyrmion field of  Eq.\ (\ref{sn}),  we obtain 
\begin{equation}
\mathcal{H}_{\rm int} = -\frac{1}{12}JS^2a^{2}\int
dxdy\left(\partial_{x}^{2}\mathbf{L}_s \cdot \partial_{x}^{2}\delta\mathbf{L} + \partial_{y}^{2}\mathbf{L}_s \cdot \partial_{y}^{2}\delta\mathbf{L}\right).
\label{LatHam-Lin}
\end{equation}
We select the quantized wave approximation for $\delta {\bf L}$:
\begin{equation}
\delta {\bf L} = \sum_{{\bf q}, \alpha} \sqrt{\frac{\hbar c^2}{2JS^2A\omega_q}}{\bf e}_{\alpha}\left(e^{-i{\bf q}\cdot {\bf r}}a_{{\bf q},\alpha} + e^{i{\bf q}\cdot {\bf r}}a^{\dagger}_{{\bf q},\alpha}\right), \label{L-quantized}
\end{equation}
with $ \alpha = x,y$. Here  $a^{\dagger}_{{\bf q},\alpha}, a_{{\bf q},\alpha}$ are operators of creation and annihilation of magnons corresponding to quantum oscillations of the N\'{e}el vector, ${\bf e}_{\alpha}$ are their polarization vectors, $A$ is the area of the xy-space, and $\omega_q = cq$. The factor under the square root in Eq. (\ref{L-quantized}) is chosen such that Eq.\ (\ref{H0}) for magnons becomes
\begin{equation}
{\cal {H}}_m = \sum_{{\bf q}, \alpha} \hbar \omega_q a^{\dagger}_{{\bf q},\alpha}a_{{\bf q},\alpha}.
\end{equation}
These two branches of antiferromagnetic magnons are quantized waves \cite{Ivanov} of $L_x$ and $L_y$ under the assumption that the N\'{e}el vector at infinity is directed along the z-axis. It is easy to see that they satisfy the conventional commutation relation for an antiferromagnet: $[L_x,L_z]=\frac{1}{2}M_z \rightarrow 0$, with $M_z$ being the magnetization. 

\begin{table*}
\begin{centering}
\begin{adjustbox}{width=17.8cm}
\begin{tabular}{|c|c|c|c|c|c|c|c|c|c|}
\hline 
n\textbackslash m & 0 & 1 & 2 & 3 & 4 & 5 & 6 & 7 & 8\tabularnewline
\hline 
0 & 0 & $0.970\times10^{-3}$ & 0 & $1.36\times10^{-3}$ & 0 & $0.659\times10^{-3}$ & 0 & $2.00\times10^{-4}$ & 0\tabularnewline
\hline 
1 & 0 & 0 & $0.284\times10^{-3}$ & 0 & $0.389\times10^{-4}$ & 0 & $1.60\times10^{-4}$ & 0 & $0.439\times10^{-4}$\tabularnewline
\hline 
2 & 0 & 0 & 0 & $0.471\times10^{-4}$ & 0 & $0.603\times10^{-4}$ & 0 & $2.25\times10^{-5}$ & 0\tabularnewline
\hline 
3 & 0 & 0 & 0 & 0 & $0.512\times10^{-5}$ & 0 & $0.604\times10^{-5}$ & 0 & $2.01\times10^{-6}$\tabularnewline
\hline 
4 & 0 & 0 & 0 & 0 & 0 & $0.389\times10^{-6}$ & 0 & $0.417\times10^{-6}$ & 0\tabularnewline
\hline 
5 & 0 & 0 & 0 & 0 & 0 & 0 & $2.17\times10^{-8}$ & 0 & $2.11\times10^{-8}$\tabularnewline
\hline 
6 & 0 & 0 & 0 & 0 & 0 & 0 & 0 & $0.941\times10^{-9}$ & 0\tabularnewline
\hline 
7 & 0 & 0 & 0 & 0 & 0 & 0 & 0 & 0 & $0.328\times10^{-10}$\tabularnewline
\hline 
\end{tabular}
\end{adjustbox}
\par\end{centering}
\caption{Transition rates in units $J/\hbar$ from state $m$ to state $n$ for $S=1/2$ and $l=1000$.}
\label{table2}
\end{table*}
The rate of the transition from the state $\psi_{m}(\lambda)$ with zero magnons to the lower
energy state $\psi_{n}(\lambda)$ ($n<m$) and one magnon is given by the Fermi golden rule \cite{Messiah},
\begin{eqnarray}
\Gamma &=& \frac{2\pi}{\hbar}\sum_{i\neq j}\langle i|\hat{H}_{\rm int}|j\rangle\langle j|\hat{H}_{\rm int}|i\rangle\delta(E_{i}-E_{j})  \\
&=& \frac{2\pi}{\hbar}\sum_{{\bf q},\alpha}|\langle\psi_{m}(\lambda)1_{{\bf q},\alpha}|\hat{H}_{\rm int}|\psi_{n}(\lambda)0\rangle|^{2}\delta(\hbar\omega_{q}-\Delta_{mn}),\nonumber \label{Gamma-gen}
\end{eqnarray}
where $\Delta_{mn}=E_{m}-E_{n}$. Substituting here $\hat{H}_{\rm int}$ of Eq.\ (\ref{LatHam-Lin}) one obtains
\begin{eqnarray}
&& \Gamma_{mn}=\frac{2\pi}{\hbar}\left(\frac{1}{12}JS^{2}a^{2}\right)^{2}\sum_{{\bf q},\alpha}|\intop d\lambda\psi_{m}^{*}(\lambda)\psi_{n}(\lambda) \times \nonumber \\
&&\int dxdy\,\left(\partial_{x}^{2}\mathbf{L}_{s}\cdot\langle1_{{\bf q},\alpha}|\partial_{x}^{2}\delta\mathbf{L_{{\bf q},\alpha}|}0_{m}\rangle +  \right. \nonumber \\ 
&& \left. \partial_{y}^{2}\mathbf{L}_{s}\cdot\langle1_{{\bf q},\alpha}|\partial_{y}^{2}\delta\mathbf{L_{{\bf q},\alpha}|}0_{m}\rangle\right)|^{2}\delta(\hbar\omega_{q}-\Delta_{mn}).
\end{eqnarray}
In this expression 
\begin{equation}
\partial_{x}^{2}\delta\mathbf{L}_{{\bf q},\alpha}=-\sqrt{\frac{\hbar c^{2}}{2JS^{2}A\omega_{q}}}{\bf e}_{\alpha}q_{x}^{2}\left(e^{-i{\bf q}\cdot{\bf r}}a_{{\bf q},\alpha}+e^{i{\bf q}\cdot{\bf r}}a_{{\bf q},\alpha}^{\dagger}\right)
\end{equation}
and thus 
\begin{equation}
\langle1_{{\bf q},\alpha}|\partial_{x}^{2}\delta\mathbf{L|}0_{m}\rangle=-\sqrt{\frac{\hbar c^{2}}{2JS^{2}A\omega_{q}}}{\bf e}_{\alpha}q_{x}^{2}e^{i{\bf q}\cdot{\bf r}}
\end{equation}
and similar for $\partial_{y}^{2}\delta\mathbf{L}$. This gives 
\begin{eqnarray}
&& \Gamma_{mn}  =  \frac{2\pi}{\hbar}\left(\frac{1}{12}JS^{2}a^{2}\right)^{2}\frac{\hbar c^{2}}{2JS^{2}A}\sum_{{\bf q},\alpha}\left|\intop d\lambda\psi_{m}^{*}(\lambda)\psi_{n}(\lambda) \right.\nonumber \\
& &  \left.  \times \int dxdy\,\left(\partial_{x}^{2}\mathbf{L}_{s}\cdot{\bf e}_{\alpha}\frac{q_{x}^{2}}{\omega_{q}}e^{i{\bf q}\cdot{\bf r}}+\partial_{y}^{2}\mathbf{L}_{s}\cdot{\bf e}_{\alpha}\frac{q_{y}^{2}}{\omega_{q}}e^{i{\bf q}\cdot{\bf r}}\right)\right|^{2}  \nonumber \\
&& \times \delta(\hbar\omega_{q}-\Delta_{mn}), \nonumber
\end{eqnarray}
which can be represented as
\begin{equation}
\Gamma_{mn}=\frac{\pi JS^2a^4 c^{2}}{144A}\sum_{{\bf q}}\frac{1}{\omega_{q}}\left(M_{x}+M_{y}\right)\delta(\hbar\omega_{q}-\Delta_{mn}), \nonumber
\end{equation}
with
\begin{equation}
M_{\alpha}\equiv\left|\intop d\lambda\psi_{n}^{*}(\lambda)\psi_{m}(\lambda)F_{\alpha}\right|^{2},\qquad\alpha=x,y,
\end{equation}
where
\begin{eqnarray}
&& F_{\alpha}  \equiv  \int dxdy\left(\partial_{x}^{2}L_{s\alpha}q_{x}^{2}+\partial_{y}^{2}L_{s\alpha}q_{y}^{2}\right)e^{i{\bf q}\cdot{\bf r}}  \nonumber \\
&& = -\left(q_{x}^{4}+q_{y}^{4}\right)\int dxdyL_{s\alpha}e^{i{\bf q}\cdot{\bf r}}=-\left(q_{x}^{4}+q_{y}^{4}\right)\tilde{L}_{s\alpha}\left(\lambda,\mathbf{q}\right). \nonumber
\end{eqnarray}

Further simplification requires computation of the Fourier transform of the skyrmion field
\begin{equation}
\tilde{L}_{s\alpha}\left(\lambda,\mathbf{q}\right)\equiv\int dxdyL_{s\alpha}\left(\lambda,\mathbf{r}\right)e^{i{\bf q}\cdot{\bf r}}.
\end{equation}
In terms of the latter
\begin{equation}
\Gamma_{mn}=\frac{\pi JS^2a^4 c^{2}}{144A}\sum_{{\bf q}}\frac{\left(q_{x}^{4}+q_{y}^{4}\right)^{2}}{\omega_{q}}\left(\tilde{M}_{x}+\tilde{M}_{y}\right)\delta(\hbar\omega_{q}-\Delta_{mn}),\label{Gamma_via_M} \nonumber
\end{equation}
where 
\begin{equation}
\tilde{M}_{\alpha}\equiv\left|\intop d\lambda\psi_{m}^{*}(\lambda)\psi_{n}(\lambda)\tilde{L}_{s\alpha}\left(\lambda,\mathbf{q}\right)\right|^{2},\qquad\alpha=x,y. \nonumber
\end{equation}
By symmetry, $\tilde{M}_{x}$ and $\tilde{M}_{y}$ make equal contributions to the rate, which simplifies to
\begin{eqnarray}
\Gamma_{mn} 
 & = & \frac{JS^{2}a^{4}}{144\hbar}\int_{0}^{\infty}dq\int_{0}^{2\pi}d\varphi_{q}q^{8}\left(\sin^{4}\varphi_{q}+\cos^{4}\varphi_{q}\right)^{2} \nonumber \\
 & \times& \tilde{M}_{x}\delta\left(q-\frac{\Delta_{mn}}{\hbar c}\right).\label{Gamma_via_M_1}
\end{eqnarray}

To calculate $\tilde{M}_{x}$, we need the in-plane component of the skyrmion field,
\begin{equation}
L_{sx}(r,\varphi_{r})=L_{\bot}(r)\sin\left(\varphi_{r}+\gamma\right).
\end{equation}
From Eq.\ (\ref{sn}) one has $L_{\bot}(r)={2\lambda r}/({\lambda^{2}+r^{2}})$. Then
\begin{eqnarray}
&& \tilde{L}_{sx}(q,\varphi_{q})  = \intop_{0}^{\infty}rL_{\bot}(r)dr\intop_{0}^{2\pi}d\varphi_{r}\sin\left(\varphi_{r}+\gamma\right) \nonumber \\  
&&\times e^{-iqr\cos\left(\varphi_{r}-\varphi_{q}\right)} = -4\pi i\sin\left(\varphi_{q}+\gamma\right)\lambda |\lambda|K_{1}(|\lambda| q). \nonumber 
\end{eqnarray}
and
\begin{equation}
\tilde{M}_{x}\equiv\left(4\pi\right)^{2}\sin^{2}\left(\varphi_{q}+\gamma\right)\left|F_{mn}(q)\right|^{2},
\end{equation}
where
\begin{equation}
F_{mn}(q)\equiv\intop d\lambda\psi_{m}^{*}(\lambda)\psi_{n}(\lambda)\lambda |\lambda|K_{1}(|\lambda| q) \label{F_nm_def}
\end{equation}
with $K_1$ being the modified Bessel function. The transition rate $\Gamma_{mn}$ of Eq.
(\ref{Gamma_via_M_1}) becomes
\begin{eqnarray}
&& \Gamma_{mn}=\frac{JS^{2}a^{4}}{144\hbar}\int_{0}^{\infty}dq\int_{0}^{2\pi}d\varphi_{q}q^{8}\left(\sin^{4}\varphi_{q}+\cos^{4}\varphi_{q}\right)^{2} \nonumber \\
&& \times \left(4\pi\right)^{2}\sin^{2}\left(\varphi_{q}+\gamma\right)\left|F_{mn}(q)\right|^{2}\delta\left(q-\frac{\Delta_{mn}}{\hbar c}\right)
\end{eqnarray}
The integral over $\varphi_{q}$ equals $19\pi/32$ independently of $\gamma$, yielding the final compact expression for the rate:
\begin{equation}
\Gamma_{mn}=\frac{19\pi^{3}JS^{2}}{288\hbar}\left(q_{mn}a\right)^{4}\left|q_{mn}^{2}F_{mn}(q_{mn})\right|^{2}, \label{nm-rates}
\end{equation}
with $q_{mn}\equiv{\Delta_{mn}}/({\hbar c})$ and  $F_{mn}$ defined by Eq. (\ref{F_nm_def}). Computation of the transition rates reduces, therefore, to the computation of the coefficients $F_{nm}$ with the wave functions of the stationary states of the skyrmion found in Section \ref{QH}. 

Since energies of skyrmion eigenstates, $E_n$, decrease very fast on increasing $n$ (see Fig.\ \ref{Bohr}) so do distances between the levels $\Delta_{mn}$ for large $m$ and $n$. In this case the smallness of  $|\lambda| q_{mn}= |\lambda| \Delta_{mn}/({\hbar c})$ allows one to replace $K_1(|\lambda | q)$ in Eq.\ (\ref{F_nm_def}) with its asymptotic form $1/(|\lambda | q)$, leading to $F_{mn}= \lambda_{mn}/q$ with $\lambda_{mn} = \int d\lambda \, \lambda \psi^*_m(\lambda)\psi_n(\lambda)$ and to 
\begin{equation}
\Gamma_{mn}=\frac{19\pi^{3}JS^{2}}{288\hbar}\left(q_{mn}a\right)^{4}\left|q_{mn}\lambda_{mn}\right|^{2}, \quad m,n \gg 1. 
\end{equation}
This makes the rates of transitions between high excited levels proportional to $\Delta_{mn}^6$ and progressively small with increasing $m$ and $n$.  Numerically obtained transition rates for $S = 1/2$ are shown in Table \ref{table2}. The selection rule related to the parity of the skyrmion states is apparent.

\section{Discussion} \label{Discussion}

We have studied quantum states of antiferromagnetic skyrmions by quantizing the scaling parameter of Belavin-Polyakov model. Our results suggest that energies and sizes of nanoscale skyrmions are quantized. Quantum mechanics must also slow down the collapse of small skyrmions, making them more stable as was proposed earlier \cite{Roldan2015}. The question is whether quantum states of skyrmions and transitions between them can be observed in experiment. 

Evidence of skyrmions has been reported in parental compounds of high-temperature superconductors \cite{exp}. In fact, many of these materials can be ideal systems for application of our theory as they consist of weakly interacting 2D antiferromagnetic layers of spins $1/2$ in a square lattice. The antiferromagnetic superexchange interaction  \cite{La} of copper spins via oxygen in CuO layers of La$_2$CuO$_4$ is $140$meV. This places lifetimes of the low-lying excited states of the skyrmion in the picosecond to nanosecond range. Transitions between upper excited states have much lower probability. 

It should be emphasized that our model is not catching the final stage of the decay of the skyrmion accompanied by the disappearance of its topological charge when it decreases to the atomic size. This requires a different mechanism. In classical mechanics on the lattice, it must occur via instability of the skyrmion shape that breaks its radial symmetry. Quantum mechanics efecively switches the skyrmion model from 2 to 2+1 dimensions where topological charge is not conserved but working it out may require more than one degree of freedom. We have not attempted to solve this problem here. 

Quantization of the quasiclassical states of skyrmions that are large compared to the lattice spacing must be captured by our model with reasonable accuracy. For such skyrmions the distance between adjacent energy levels is of order $\Delta \sim |E| \sim 2\pi JS^2a^2/(3\lambda ^2)$. Transition from classical to quantum dynamics should occur at temperatures satisfying $T \lesssim \Delta$, that is, for sizes satisfying  $ \lambda/a \lesssim \sqrt{2\pi JS^2/(3T)}$. For La$_2$CuO$_4$ this gives $ \lambda \lesssim 3a$ at room temperature and $ \lambda \lesssim 30a$ at helium temperature. The latter makes observation of quantum behavior of skyrmions promising at low temperature. If the collapse of a skyrmion could somehow be visualized with the help of modern imaging techniques its quantum nature would reveal itself in a large jump from a nanometer size to the atomic size. 

At elevated temperatures skyrmions would be created by thermal fluctuations. When many small skyrmions are present, quantization of their energy levels must lead to the peaks in the absorption and noise spectra corresponding to the transitions between the levels. If  low temperatures were required for a precision study  of quantization of skyrmion states a practical question would be how to create a sufficient number of antiferromagnetic skyrmions at such temperatures. One way to do it could be via rapid cooling of the sample accompanied by quantum relaxometry \cite{Finco2021} of the antiferromagnetic state that settles in. 

In this article we have not addressed quantum mechanics of ferromagnetic skyrmions. In the absence of interactions other than the Heisenberg exchange, the ferromagnetic dynamics, unlike antiferromagnetic dynamics, is massless. Consequently, the quantization method we applied, cannot be easily extended to a ferromagnet. This already showed up in our study of the classical collapse \cite{CCG-PRB2012}. The collapse of the antiferromagnetic skyrmion was investigated by two methods, via dynamical equation for $\lambda(t)$ and by solving numerically the full two-sublattice Heisenberg spin model with the initial state containing a skyrmion. Excellent agreement between the two methods was achieved. On the contrary, the study of the collapse of the ferromagnetic skyrmion relied on the second method only. The observed dynamics was very different, with the lifetime of the ferromagnetic skyrmion scaling as $[\hbar/(JS^2)](\lambda_0/a)^5$ compared to $[\hbar/(JS^2)](\lambda_0/a)^2$ for the antiferromagnetic skyrmion. Due to its complexity the full spin model, however, is not well suited for computing energy levels of a small ferromagnetic skyrmion. Finding a method for a 2D ferromagnet similar to that for a 2D antiferromagnet, as well as inclusion in the model of other interactions such as Dzyaloshinskii-Moriya, dipole-dipole, Zeeman, magnetic anisotropy, remains a challenging task.\\

\section{Acknowledgements}

This work has been supported by the Grant No. DE-FG02-93ER45487 funded by the U.S. Department of Energy, Office of Science.

\end{document}